\documentclass[prl,twocolumn,showpacs,floatfix,amsfonts]{revtex4}
\usepackage{graphicx,graphics,color,epsfig}
\usepackage{bm}
\usepackage{amsmath}
\usepackage{amssymb}
\usepackage{epstopdf}
\begin{document}
\preprint{}
\title{Exciton Condensate Modulation  in Electron-Hole Bilayers: A Real-Space Visualization}
\author{Jian-Xin Zhu and A. R. Bishop}
\affiliation{Los Alamos National Laboratory,
Los Alamos, New Mexico 87545, USA}
\date{\today}
\begin{abstract}
We study the texture of the exciton condensate at low temperatures in an independently gated  electron-hole bilayer system.  A model Hamiltonian is solved in real space within a mean-field approximation. It is found that, with increased electron-hole density polarization, the system experiences phase transformations from the zero center-of-mass momentum superfluid state, through one- and two-dimensional exciton pair modulated states, into the normal state. At weak density polarization, the modulating state resembles the Larkin-Ovchinikov state in superconductors in the presence of an exchange field in the weak-coupling BCS limit, and becomes stripe-like in the strong coupling BEC limit.  In the one-dimensional modulated phase, the density of states exhibits low-energy intra-gap resonance quasiparticle states, which are localized in the nodal region.
\end{abstract}
\pacs{73.21,Fg, 71.35.Lk, 71.10.Li}
\maketitle

\narrowtext

The Fulde-Ferrel-Larkin-Ovchinnikov (FFLO) state~\cite{PFulde:1964,AILarkin:1964} represents  a class of unconventional superconducting states, where the superconducting order parameter  modulates in {\em  real space}. It is {\em dual} to the other more familiar class of unconventional superconducting states, in which the superconducting order parameter modulates (or even changes sign) in {\em momentum space}.  In the FFLO state, the Cooper pairs form between two electrons with  $(\mathbf{k}+\mathbf{q}/2,\uparrow)$ and $(-\mathbf{k}+\mathbf{q}/2,\downarrow)$ of momentum and spin configuration. The center-of-mass momentum $\mathbf{q}$ is dependent on the extent to which the Fermi surface is Zeeman split by an exchange field.   Although the FFLO state was predicted more than four decades ago, an undisputed verification of this state in superconductors remains a great experimental challenge.  A major reason lies in the fact that, in most superconductors, the Pauli paramagnetic effect of the applied magnetic field is negligibly small  compared with the orbital breaking effect. Owing to recent advances in the discovery of new materials and new technology, there is a revival of interest in the FFLO-type modulated states in a wide range of systems. Among them, newly discovered  heavy fermion superconductors~\cite{ABianchi:2003,HARadovan:2003,YMatsuda:2007}, low-dimensional organic 
superconductors~\cite{JSingleton:2000,MATanatar:2002,SUji:2006,IJLee:1997,JShinagawa:2007}, and trapped cold Fermion atoms~\cite{MWZwierlein:2006,
GBPartridge:2006,YShin:2006,DESheehy:2006,MMForbes:2005,ABulgac:2006} are good candidates for the emergence of a superconducting FFLO state. In parallel,  semiconductor bilayer systems~\cite{JPEisenstein:2004} represent another promising context  for the concept of FFLO-like states and the fundamental physics of the BCS-BEC crossover, in which there is  condensation of  electron and hole pairs. 

In electron-hole bilayers, the density of the electrons and holes can be varied independently of each other. The effect of electron-hole density imbalance here resembles the exchange effect of an applied magnetic field in a superconductor. Such an effect on the BCS-BEC crossover in semiconductor electron-hole bilayers has  recently been investigated at zero temperature by Pieri and co-workers~\cite{PPieri:2007}, where an instability toward to the FFLO phase  was identified. A more recent study~\cite{KYamashita:2009}, by treating the FFLO phase and the Sarma phase~\cite{GSarma:1963} on an equal footing, indicated that the FFLO state can be well stabilized by order parameter mixing effect.  We note that,  though collectively known as the FFLO state, the form proposed by Fulde and Ferrel (FF)~\cite{PFulde:1964} and that by Larkin and  Ovchinnikov (LO)~\cite{AILarkin:1964} are slightly different. In the FF form, the order parameter has a homogeneous magnitude but with a modulated complex phase factor, that is a single-wavevector plane wave; while, in the LO form, the order parameter is real and spatially modulated. In Ref.~\cite{PPieri:2007,KYamashita:2009}, the authors solve the problem directly in momentum space, which limits consideration to the FF form. In this paper, we consider a tight-binding model for an electron-hole bilayer and solve the BCS-type equation self-consistently in real space. It is found that the LO form is always stabilized. Of particular interest, we find a crossover from the LO state to an exciton stripe state as the pairing interaction is tuned from the BCS limit into the the BEC limit. Our local density of states calculations show the existence of intra-gap states,  which are localized on the nodal lines, in the one-dimensional LO phase.

\begin{figure}[h]
\includegraphics[width=0.8\linewidth]{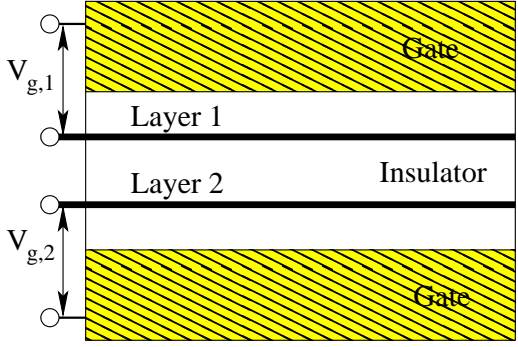}
\caption{(Color) Schematic of a bilayer system. The carrier in each layer has electron or hole nature, which is determined by the individual band structure. The chemical potential in each layer is adjusted by the respective gate voltage.}
\label{FIG:setup}
\end{figure}
 
We start with a generic two-layer system,  schematically shown in Fig.~\ref{FIG:setup}. The two layers are separated by an insulator such that the interlayer tunneling is negligible. The chemical potential in each layer can be tuned by gate voltages $V_{g,+}$ and $V_{g,-}$. In the tight-binding model for a square lattice in each layer, the Hamiltonian can be written as:
\begin{equation}
\mathcal{H} = \sum_{ij,\sigma} [-t_{ij,\sigma} -\mu_{\sigma}\delta_{ij}]c_{i\sigma}^{\dagger}c_{j\sigma} + \frac{1}{2} \sum_{ij,\sigma} g_{ij} n_{i,\sigma}n_{j,\bar{\sigma}} \;.
\label{EQ:Hamil1}
\end{equation}
Here $c_{i\sigma}^{\dagger}$ ($c_{i\sigma}$) are the creation (annihilation) operators of an electron  at site $i$ in layer $\sigma=1$ or $2$. The quantities $t_{ij,\sigma}$ and $\mu_{\sigma}=V_{g,\sigma}$ are the hopping integrals and chemical potentials in each individual layer, respectively. We introduce $g_{ij}$ as the inter-layer particle-particle  coupling strength. In Hamiltonian~\ref{EQ:Hamil1}, the particle-particle interaction within each individual layer is ignored, since they are expected to mainly  renormalize each band. In our setting, explicit spin quantum numbers are omitted. As such, the index $\sigma$ can be regarded as a pseudospin index, which we will use interchangebly, $\sigma= +(-)$ or $\uparrow(\downarrow)$. 
In the mean-field approximation, the above Hamiltonian can be written as 
\begin{eqnarray}
\mathcal{H}_{\text{MF}} &=& \sum_{ij,\sigma} [-t_{ij,\sigma} -\mu_{\sigma}\delta_{ij}]c_{i\sigma}^{\dagger}c_{j\sigma}  + \sum_{ij} [\Delta_{ij} c_{i1}^{\dagger} c_{j2} + h.c.]  \nonumber \\
&& + \sum_{ij} \vert \Delta_{ij} \vert^{2}/g_{ij}\;,
\label{EQ:Hamil2}
\end{eqnarray}
where the order parameter is defined as $\Delta_{ij} = g_{ij} \langle c_{i1}c_{j2}^{\dagger}\rangle$.
Through a canonical transformation, the above Hamiltonian can be diagonalized by solving the following Bogoliubov-de Gennes (BdG) equation:
\begin{equation}
\sum_{j} \left( \begin{array}{cc} 
h_{ij,1} & \Delta_{ij} \\
\Delta_{ji}^{*} & h_{ij,2} 
\end{array} \right) 
\left( 
\begin{array}{c}
u_{j}^{n} \\ v_{j}^{n} 
\end{array} \right)
=E_{n} \left( \begin{array}{c} u_{i}^{n} \\ v_{i}^{n} \end{array} 
\right) \;,
\label{EQ:BdG}
\end{equation}
where the single-particle Hamiltonian and the pair potential are given by 
\begin{equation}
h_{ij,\sigma} = -t_{ij,\sigma} -\mu_{\sigma} \delta_{ij}\;,
\end{equation}
and 
\begin{equation}
\Delta_{ij} = \frac{1}{2} \sum_{n} u_{i}^{n} v_{j}^{n*}\tanh\left( \frac{E_{n}}{2k_{B}T}\right)\;,
\end{equation}
and $(u_{i}^n,v_{i}^n)^{\text{\small Transpose}}$ are the eigenvector corresponding to the eigenenergy $E_{n}$. The quasiparticle energy is measured with respect to the Fermi energy.

\begin{figure}[h]
\includegraphics[width=0.8\linewidth]{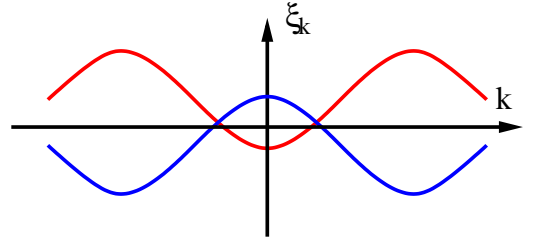}
\caption{(Color) Schematic of the energy dispersion for electron (red) and hole (blue) layers.}
\label{FIG:disp}
\end{figure}

The above formalism is generic and applicable to both electron-electron or electron-hole bilayer systems, as determined by the band structure within each layer.  
Within the nearest-neighbor hopping approximation, the band dispersion for the clean case can be written as 
\begin{equation}
\xi_{k\sigma} = -2t_{\sigma} (\cos k_{x} + \cos k_{y}) - \mu_{\sigma}\;,
\end{equation}
with $t_{1}>0$ and $t_{2}>0$  for the electron-electron bilayers, and $t_{1} >0$ and $t_{2}<0$ for the electron-hole bilayers. A schematic of normal-state band dispersion is shown in Fig.~\ref{FIG:disp}.
Also the particle-particle interaction is long-ranged when the screening effect is weak, as considered in usual coupled quantum wells, and short-ranged when the screening effect is strong, which might be relevant when the metallic gate structures are present in the vicinity of quantum wells. For the purpose of the present discussion, we consider the electron-hole bilayer system, and consider a short-ranged Coulomb interaction with $g_{ij}=g\delta_{ij}$ such that $\Delta_{ij}=\Delta_{i} \delta_{ij}$.

\begin{figure}[b]
\includegraphics[width=1\linewidth]{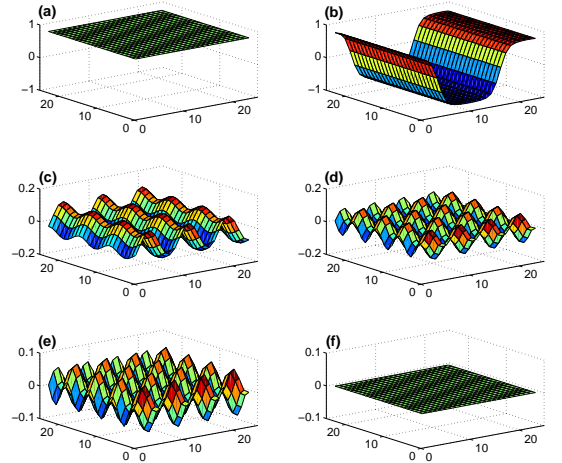}
\caption{ (Color) The spatial dependence of the exciton order parameter for changing $h=0$ (a), $0.2$ (b), $0.6$ (c), $0.7$ (d), $1.0$ (e) and $1.60$. The temperature is chosen as $T=10^{-3}$. 
The average chemical potential is chosen as $\mu=-0.5$.}
\label{FIG:OP}
\end{figure} 

We solve the above set of BdG equations self-consistently via exact diagonalization.  Within the above mean-field treatment,  the quasiparticle energy dispersion in the ``quenched'' uniform exciton condensate, that is, $\Delta_i = \Delta_0$, is of the form:
\begin{equation}
E_{\pm}(\mathbf{k}) = \frac{\xi_{\mathbf{k},1} + \xi_{\mathbf{k},2}}{2} \pm \sqrt{ \left( \frac{\xi_{\mathbf{k},1} - \xi_{\mathbf{k},2}}{2} \right)^2 + \vert \Delta_0\vert^{2} }\;.
\end{equation} 
From this energy dispersion form, one can see that the role of an exchange field in superconductors is played by the chemical potential $\mu_{1,2}$. However, the {\em independent} tunability of the hopping integral $t_{1,2}$ (or the effective bandwidth) within each layer has no counterpart in the case of superconductors.  We note that the parameters $t_{1,2}$ are related to the effective mass $m_{1,2}$ appearing in the continuum limit~\cite{PPieri:2007,KYamashita:2009}. 
When $\xi_{\mathbf{k},1} = -\xi_{\mathbf{k},2}$, the first term  on the RHS  disappears, while the first term in 
the square is maximized. The quasiparticles are fully gapped, suggesting the ``quenched'' uniform state with zero momentum pairing is a truly favored state. However, when $\xi_{\mathbf{k},1} \neq - \xi_{\mathbf{k},2}$, there emerges an instability toward a {\em finite}-momentum pairing state. 
This analysis is consistent with that given in Refs.~\onlinecite{PPieri:2007,KYamashita:2009}, and will be corroborated by the numerical calculations presented below.

In our numerical calculations, we choose to measure all energy quantities in units of the hopping integral
for the electron layer and set $t_{1}=1$.  The Boltzmann constant is set to $k_{B}=1$. A typical system size of $N_{L}= 24 \times 24 $ is used throughout the calculation and the convergence criterion, set as the difference between two consecutive iterations, is no larger than $3\times 10^{-5}$. For the convenience of discussion, we introduce the average chemical potential and pseudospin polarization potential as $\mu=(\mu_{1}-\mu_{2})/2$ and $h=(\mu_{1}+\mu_{2})/2$, and the degree of bandwidth mismatch $\alpha=-t_{2}/t_{1}$.  Without loss of generality, we take $\alpha =1$ throughout this work.

\begin{figure}[b]
\includegraphics[width=0.7\linewidth]{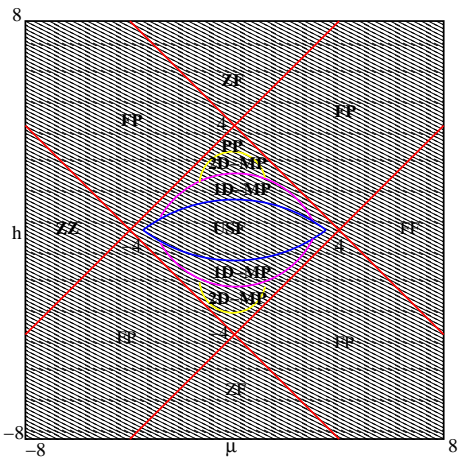}
\caption{(Color) Schematic of the electron-hole bilayer phase diagram, covering three superfluid phases (USF, 1D-MP, and 2D-MP), and several normal phases (ZZ, FF, ZF, FP, PP).}
\label{FIG:phase}
\end{figure}

In Fig.~\ref{FIG:OP}, we show the order parameter  texture on a square lattice for varying polarization field strength, $h$,  at fixed values of $\mu=-0.50$ and $g_0=3$. When the polarization degree of freedom $h=0$, the exciton order parameter is uniform (see Fig.~\ref{FIG:OP}(a)). As $h$ deviates from zero, the order parameter first exhibits a one-dimensional modulation (see Fig.~\ref{FIG:OP}(b)).  In this regime, the order parameter changes sign across the nodal lines. Our numerical calculations  demonstrates (not shown) that the sign-change slope becomes steeper as the pairing strength is increased such that the nodal lines are replaced by sharp domain walls --- a manifestation of a BCS-BEC crossover.  With increased polarization, the modulation periodicity is shortened, and the one-dimensional modulation gives way to a two-dimensional modulation (see Fig.~\ref{FIG:OP}(c-e)). Simultaneously,  
 the amplitude of the order parameter is decreased. Ultimately, the system enters a polarized normal state with even larger $h$ (see Fig.~\ref{FIG:OP}(f)). Our calculations find that the LO-like state, which breaks the translational symmetry of the underlying lattice, is always favorable. 

In Fig.~\ref{FIG:phase}, we provide a schematic of the typical  $\mu$-$h$ phase diagram for the electron-hole bilayer defined on a square lattice. The phase diagram is constructed with delineating lines $\mu+h = \pm 4$ and $-\mu +h =\pm 4$. For $\mu + h < -4$, there are no electrons occupying the electron band while for $ \mu + h >4$, the electron band is fully occupied by electrons.  Similarly, 
for $-\mu + h <-4$, the hole band is fully occupied, while for $-\mu + h > 4$ the hole band is empty.
As such,  we can map out the normal state phases, including a zero occupation phase (ZZ),  full occupation phase (FF),  a phase with one of the layers fully occupied with the other empty (ZF), a fully polarized phase with partial occupation of only one type of  carrier in the respective layer (FP), and a partially polarized phase (PP).  The interesting exciton phases are limited in the area bound by the aforementioned four delineating lines. They are categorized into a BCS (or BEC) uniform superfluid (USF) phase, and one- (1D) and two-dimensional (2D) modulated phases (MP). 

\begin{figure}[t]
\includegraphics[width=0.9\linewidth]{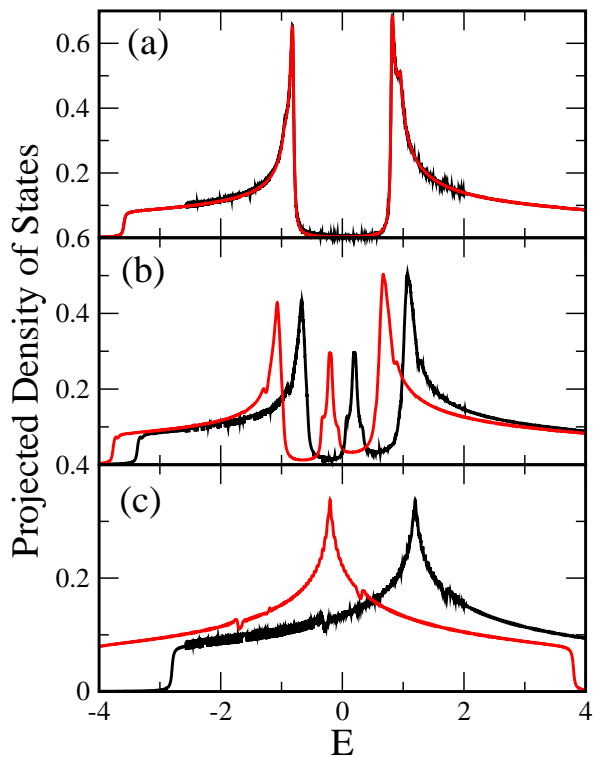}
\caption{ (Color) Average projected density of states for electron (black) and hole (red) layers at  various values of $h=0$ (a), 0.2 (b), and 0.7 (c). The average chemical potential $\mu=-0.5$.}
\label{FIG:DOS}
\end{figure}

\begin{figure}[h]
\includegraphics[width=1.0\linewidth]{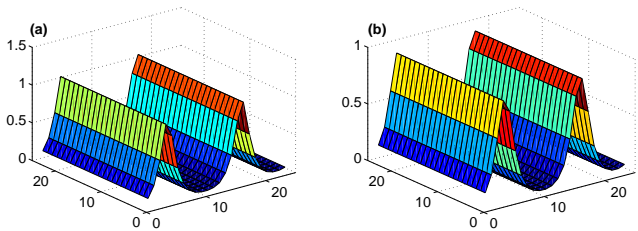}
\caption{ (Color) Spatial variation of the local density of states at $E=0.184$ (a) and -0.190 (b), corresponding to the intra-gap peaks shown in Fig.~\ref{FIG:DOS}(b).}
\label{FIG:Image}
\end{figure}

The inhomogeneous exciton state has physical implications. We calculate the average local density of states for the electron and hole layers:
\begin{subequations} 
\begin{eqnarray}
\bar{\rho}_{e}(E) &=& \frac{1}{N_{L}}\sum_{i,n} \vert u_{i}^{n}\vert^{2} \delta(E-E_n) \;, \\
\bar{\rho}_{h}(E) &=& \frac{1}{N_{L}}\sum_{i,n} \vert v_{i}^{n}\vert^{2} \delta(E+E_n) \;.
\end{eqnarray}
\end{subequations}
The Dirac $\delta$ function in the above equations is approximated by a Lorenzian 
$\delta(E-E_n) \rightarrow (\gamma/\pi)[(E-E_{n})^2+\gamma^2]^{-1}$ with $\gamma=0.02$.
In the numerical calculation, we have also used a supercell technique to partially remove the finite size effect.   In Fig.~\ref{FIG:DOS}, we show the average local density of states for various values of the polarization strength $h$. When $h=0$ (see Fig.~\ref{FIG:DOS}(a)), the LDOS exhibits a standard BCS-like quasiparticle gap around zero energy, since the exciton condensate is uniform (corresponding to Fig.~\ref{FIG:OP}(a)).  In this phase, the quasiparticle density of states are identical for the electron and hole layers. When $h$ is small but finite, for which the system is in the 1D exciton modulated state (corresponding to Fig.~\ref{FIG:OP}(b)), the density of states exhibits resonance peaks within the BCS-like gap (see Fig.~\ref{FIG:DOS}(b)). An equal number of such resonance peaks appear in the density of states for the electron and hole layers, respectively. We notice that in the BEC limit, multiple peaks can appear in the projected density of states. In addition, since  
the density is imbalanced (for non-zero $h$),  the overall density of states spectrum 
is shifted in opposite directions 
for the electron and hole layers.  When $h$ is further increased (see Fig.~\ref{FIG:DOS}(c)), the shift of the overall density of states spectrum is increased for the electron and hole layers, and the quasiparticle gap-structure becomes invisible because the  exciton order parameter has been significantly suppressed (corresponding to Fig.~\ref{FIG:OP}(d)). 

To understand the nature of the quasiparticle peaks exhibited in the density of states for the 1D modulated state, we present in Fig.~\ref{FIG:Image} the spatial variation of the local density of states at the resonance energies. It can be seen that the local density of states at these energies has a maximum at the nodal lines and shows an exponential-like decaying behavior, indicating that these states are  bound states.  

In conclusion, we have studied the texture of the exciton condensate at low temperatures in two parallel and independently gated  electron-hole layers by solving a mean-field microscopic model Hamiltonian in real space.  We have found that, with increased density imbalance, the system experiences  transformations from the zero center-of-mass momentum superfluid state, through one- and two-dimensional exciton pair modulated states, into the normal state. At weak density polarization, the modulated state resembles the Larkin-Ovchinikov state in superconductors in the presence of an exchange field in the weak-coupling BCS limit, and becomes stripe-like in  the strong coupling BEC limit.  In this phase, the density of states exhibits low-energy intra-gap resonance peaks. A local density of states imaging analysis indicates that these peaks correspond to bound state trapped near the nodal lines.  Since the local density of states is proportional to the differential tunneling conductance as measured by scanning tunneling spectroscopy/microscopy, the inhomogeneous exciton modulated state should be experimentally accessible~\cite{AVBalatsky:2006}. Therefore, an electron-hole bilayer system with density imbalance controlled by gate voltage, will provide an alternative setting for testing exotic phases proposed in superconductors.

We acknowledge the support of the
National Nuclear Security Administration
of the U.S. DOE  at the Los Alamos National Laboratory
under Contract No. DE-AC52-06NA25396, the U.S. DOE Office of Science, and the LDRD Program at LANL.


\begin{thebibliography}{99}

\bibitem{PFulde:1964} P. Fulde and R. A. Ferrell, Phys. Rev. {\bf 135}, A550 (1964).

\bibitem{AILarkin:1964} A. I. Larkin and Yu. N. Ovchinnikov, Zh. Eksp. Teor. Fiz. {\bf 47}, 1136 (1964) [Sov. Phys. JETP {\bf 20}, 762 (1969)].

\bibitem{ABianchi:2003} A. Bianchi {\em et al.}, 
Phys. Rev. Lett. {\bf 91}, 187004 (2003).

\bibitem{HARadovan:2003} H. A. Radovan {\em et al.}, 
Nature (London) {\bf 425}, 51 (2003).

\bibitem{YMatsuda:2007} For a review, see Y. Matsuda and H. Shimahara, J. Phys. Soc. Jpn. {\bf 76}, 051005 (2007).

\bibitem{JSingleton:2000} J. Singleton {\em et al.}, J. Phys.: Condens. Matter {\bf 12}, L641 (2000).

\bibitem{MATanatar:2002} M. A. Tanatar {\em et al.}, Phys. Rev. B {\bf 66}, 134503 (2002).

\bibitem{SUji:2006} S. Uji {\em et al.}, Phys. Rev. Lett. {\bf 97}, 157001(2006).

\bibitem{IJLee:1997} I. J. Lee {\em et al.}, Phys. Rev. Lett. {\bf 78}, 3555 (1997).

\bibitem{JShinagawa:2007} J. Shinagawa {\em et al.}, Phys. Rev. Lett. {\bf 98}, 147002 (2007).

\bibitem{MWZwierlein:2006} M. W. Zwierlein {\em et al.}, Science {\bf 311}, 492 (2006).

\bibitem{GBPartridge:2006} G. B. Partridge {\em et al.}, Science {\bf 311}, 503 (2006).

\bibitem{YShin:2006} Y. Shin {\em et al.}, Phys. Rev. Lett. {\bf 97}, 040401 (2006).

\bibitem{DESheehy:2006} D. E. Sheehy and L. Radzihovsky, Phys. Rev. Lett. {\bf 96}, 060401 (2006).

\bibitem{MMForbes:2005} M. M. Forbes, Phys. Rev. Lett. {\bf 94}, 017001 (2005).

\bibitem{ABulgac:2006} A. Bulgac, M. M. Forbes, and A. Schwenk, Phys. Rev. Lett. {\bf 97}, 020402 (2006).

\bibitem{JPEisenstein:2004} For a review, see J. P. Eisenstein and A. H. MacDonald, Nature {\bf 432}, 691 (2004).

\bibitem{PPieri:2007} P. Pieri, D. Neilson, and G. C. Strinati, Phys. Rev. B {\bf 75}, 113301 (2007).

\bibitem{KYamashita:2009} K. Yamashita, K. Asano, and T. Ohashi, arXiv:0908.2492.

\bibitem{GSarma:1963} G. Sarma, J. Phys. Chem. Sol. {\bf 24}, 1029 (1963).

\bibitem{AVBalatsky:2006} A. V. Balatsky, I. Vekhter, and J.-X. Zhu, Rev. Mod. Phys. {\bf 78}, 373 (2006).


\end{thebibliography}
\end{document}